\documentclass[showpacs,10pt,twocolumn,prb]{revtex4-1}
\usepackage{amsmath}
\usepackage{amssymb}
\usepackage{graphics}
\usepackage{epsfig}
\usepackage{CJK}
\usepackage{color}

\begin{document}

\title{Superconductivity in cubic La$_{3}$Al with interstitial anionic electrons}
\author{Zhijun Tu$^{1,2,\dag}$, Peihan Sun$^{3,\dag}$, Donghan Jia$^{4}$, Huiyang Gou$^{4}$, Kai Liu$^{1,2,*}$, and Hechang Lei$^{1,2,*}$}
\affiliation{$^{1}$School of Physics and Beijing Key Laboratory of Optoelectronic Functional Materials \& MicroNano Devices, Renmin University of China, Beijing 100872, China\\
$^{2}$Key Laboratory of Quantum State Construction and Manipulation (Ministry of Education), Renmin University of China, Beijing 100872, China\\
$^{3}$Department of Physics, School of Science, Hebei University of Science and Technology, Shijiazhuang, Hebei  050018, China\\
$^{4}$Center for High Pressure Science and Technology Advanced Research, Beijing 100193, China}
\date{\today}

\begin{abstract}

We report the observation of superconductivity in cubic La$_{3}$Al single crystal. It shows a metallic behavior at a normal state without observable structural transition and enters the superconducting state below $T_{c}\sim$ 6.32 K. Detailed characterizations and analysis indicate that cubic La$_{3}$Al is a bulk type-II BCS superconductor. Moreover, theoretical calculations show that it can host interstitial anionic electrons, which are located at the body center of cubic unit cell, and confirm the electron-phonon coupling as the superconducting mechamism. Thus, cubic La$_{3}$Al can be regarded as an novel electride superconductor.

\end{abstract}

\maketitle

\section{Introduction}

Electrides are a unique class of materials in which electrons serve as anions.\cite{Dye1,Dye2,MiaoM} These anionic electrons are mainly located in the interstitial positions rather than belong to certain anions.
Due to the loosely bound nature of such interstitial anionic electrons (IAEs), electrides can exhibit many of unique properties, such as low work functions,\cite{Toda,Lee,ZhangX} high hyperpolarizabilities,\cite{XuHL} low-temperature thermionic emission,\cite{Phillips} very strong reducibility,\cite{KimYJ} and efficient catalytic activities.\cite{Kitano,KangSH}
These exotic properties of electrides have attracted intensive interests in both fundamental science and practical applications. For example, electride can be used as a novel cathode material for organic light-emitting diodes and an efficient catalysts for ammonia synthesis.\cite{Kitano,Kim1,HuJ}

Besides the superior catalytic and electronic properties, some quantum properties like magnetism, band topology, and superconductivity have also been predicted and discovered in electrides.\cite{LeeSY,Hirayama,ZhaoZ,LiuZ1,LiuZ2,HouJ,ZhaoY,GuoZ,YouJY,HuangHM,WanZ,ZhangX2,WangQ,Shimizu,Struzhkin,Hosono1,Matsuishi,Miyakawa,ZhangY}
In particular, theoretical calculations have predicted that some of electrides at high pressure can exhibit rather high superconducting transition temperatures $T_{c}$'s, which are beyond the McMillan limit ($\sim$ 40 K) or even close to the temperature of liquid nitrogen.
For example, hexagonal Li$_{5}$N at 150 GPa and cubic Li$_{8}$Au at 250 GPa were predicted to show superconductivity with $T_{c}=$ 49 K and 73 K, respectively.\cite{WanZ,ZhangX2}
Moreover, the cubic phase of Li formed above 41 GPa (space group $I\overline{4}3d$) is regraded as an electride superconductor with  $T_{c}\sim$ 20 K.\cite{Shimizu,Struzhkin}
Thus, high-pressure electrides could be a novel platform to explore high-temperature superconductivity.
When compared to the intensively studied high-pressure electride superconductors, the experimental studies on electride superconductors at ambient pressure are very rare.
One of example is inorganic electride [Ca$_{24}$Al$_{28}$O$_{64}$]$^{4+}$$\cdot$4e$^{-}$, which exhibits superconductivity at $T_{c}\sim$ 0.4 K.\cite{Hosono1,Matsuishi,Miyakawa}
Another one is Nb$_{5}$Ir$_{3}$ with one-dimensional IAEs that has been found to become superconducting below with $T_{c}\sim$ 9.4 K.\cite{ZhangY}

Very recently, we proposed that cubic La$_3$In is a candidate of electride superconductor with $T_{c}\sim$ 9.4 K.\cite{TuZ} For this electride superconductor, La and In atoms occupy the face-centered and vertex positions of cubic structure, receptively. The IAEs are located at the body center of cubic unit cell, surrounded by six La atoms.
Actually, the cubic La$_3$In belongs to the material family with famous Cu$_{3}$Au structure (space group $Pm$-3$m$, No. 221), which includes thousands of binary compounds.
Moreover, isostructural La$_{3}$X (X = Ga, Sn, Tl) also show superconductivity with $T_{c}\sim$ 5.8 - 9.0 K.\cite{Bucher,Garde}
Interestingly, La$_3$Al has two polymorphs. One has a hexagonal structure with La kagome lattice (space group $P$6$_3$/$mmc$, No. 194).\cite{Garde,H-La3Al} Hexagonal La$_3$Al shows a superconductivity with $T_{c}=$ 5.80 -- 6.37 K.\cite{Garde, H-La3Al,Chen} Anther polymorph of La$_3$Al exhibits a cubic structure, isostructural to La$_{3}$In.\cite{PengD}
Yet, the physical properties of cubic La$_{3}$Al are still unknown.
In this work, we grew the cubic single crystals of La$_{3}$Al and studied their physical properties in detail. We discovered that cubic La$_{3}$Al shows a superconductivity with $T_{c} \sim$ 6.32 K.
Further experimental and theoretical results indicate that cubic La$_{3}$Al is an intermediately coupled type-II BCS superconductor with IAEs located at the center of empty La$_{6}$ octahedra.

\section{Experimental and calculation details}

Single crystal of La$_{3}$Al was synthesized using the self-flux method. La chunk (purity 99.9 $\%$) and Al powders (purity 99.9 $\%$) were mixed in a molar ratio of 76 : 24.
The mixture was loaded into a Nb crucible and sealed in a quartz ampoule under a partial argon atmosphere. The sealed quartz ampoule was then heated to 1223 K for 24 h and soaked there for 24 h. Subsequently, it was cooled down to 813 K at a rate of 2 K/h. Finally, the ampoule was removed from the furnace and La$_{3}$Al single crystals were separated from the flux with a centrifuge. The typical size of the La$_{3}$Al single crystal is about 0.8$\times$0.8$\times$0.6 mm$^{3}$.
The X-ray diffraction (XRD) was performed using a Bruker D8 X-ray diffractometer with Cu $K_\alpha$ radiation ($\lambda=$ 1.5418 \AA) at room temperature. Single crystal XRD patterns were collected using a Bruker D8 VENTURE PHOTO II diffractometer with multilayer mirror monochromatized Mo $K_\alpha$ ($\lambda=$ 0.71073 \AA) radiation. Unit cell refinement and data merging were done with the SAINT program, and an absorption correction was applied using Multi-Scans.
The composition of La$_{3} $Al single crystal was determined by examination of multiple points on the crystals using energy dispersive X-ray spectroscopy (EDX) in in a FEI Nano 450 scanning electron microscope. Electrical transport measurement was performed in  superconducting magnet system (Cryomagnetics, C-Mag Vari-9). Heat capacity  and magnetization measurements was carried out in Quantum Design Physical Property Measurement System(PPMS-14T) and Magnetic Property Measurement System (MPMS3), respectively.

The electronic structures of La$_3$Al were studied based on the density functional theory (DFT) calculations with the projector augmented wave (PAW) method \cite{PAW1,PAW2} as implemented in the Vienna \textit{ab initio} simulation package (VASP).\cite{VASP1,VASP2,VASP3} The generalized gradient approximation of the Perdew-Burke-Ernzerhof (PBE) type was adopted for the exchange-correlation functional.\cite{PBE} The energy cutoff of the plane-wave basis was set to 520 eV. A 16$\times$16$\times$16 Monkhorst-Pack {\bf k}-point mesh was used to sample the Brillouin zone (BZ). The Fermi surface was broadened by the Gaussian smearing method with a width of 0.05 eV.  Both the lattice parameters and the internal atomic positions were optimized. The convergence tolerances of force and energy were set to 0.01 eV/{\AA} and $10^{-5}$ eV, respectively. The maximally localized Wannier functions method \cite{MLWF1,MLWF2} was used to calculate the Fermi surface, which was visualized with the FermiSurfer package.\cite{FermiSurf}

To investigate the phonon spectra and electron-phonon coupling (EPC), the density functional perturbation theory (DFPT) \cite{DFPT1,DFPT2} calculations were performed with the Quantum ESPRESSO (QE) package.\cite{QE} The interactions between electrons and nuclei were described by the RRKJ-type ultrasoft pseudopotentials \cite{RRKJ} from the PSlibrary.\cite{PSlibrary1,PSlibrary2} The kinetic energy cutoff for the wave-functions was set to 90 Ry. The Gaussian smearing method with a width of 0.004 Ry was used for the Fermi surface broadening. In the calculations of the dynamical matrix and the EPC, the BZ was sampled with a 4$\times$4$\times$4 {\bf q}-point mesh and a 60$\times$60$\times$60 {\bf k}-point mesh, respectively. Based on the EPC theory, the Eliashberg spectral function $\alpha^2F(\omega)$ is defined as \cite{EPC}

\begin{equation}
\alpha^2F(\omega)=\frac{1}{2{\pi}N(\varepsilon_F)}\sum_{{\bf q}\nu}\delta(\omega-\omega_{{\bf q}\nu})\frac{\gamma_{{\bf q}\nu}}{\hbar\omega_{{\bf q}\nu}},
\end{equation}
where $N(\varepsilon_F)$ is the density of states (DOS) at Fermi level $\varepsilon_F$, $\omega_{{\bf q}\nu}$ is the frequency of the $\nu$-th phonon mode at wave vector {\bf q}, and $\gamma_{{\bf q}\nu}$ is the phonon linewidth,\cite{EPC}

\begin{equation}
\gamma_{{\bf q}\nu}=2\pi\omega_{{\bf q}\nu}\sum_{{\bf k}nn'}|g_{{\bf k+q}n',{\bf k}n}^{{\bf q}\nu}|^2\delta(\varepsilon_{{\bf k}n}-\varepsilon_F)\delta(\varepsilon_{{\bf k+q}n'}-\varepsilon_F),
\end{equation}
in which $g_{{\bf k+q}n',{\bf k}n}^{{\bf q}\nu}$ is the electron-phonon coupling matrix element. The total electron-phonon coupling constant $\lambda$ can be obtained via \cite{EPC}

\begin{equation}
\lambda=\sum_{{\bf q}\nu}\lambda_{{\bf q}\nu}=2\int{\frac{\alpha^2F(\omega)}{\omega}d\omega}.
\end{equation}

The superconducting transition temperature $T_c$ can be determined by substituting the EPC constant $\lambda$ into the McMillan-Allen-Dynes formula,\cite{Tc-formula1}

\begin{equation}\label{eq5}
T_c=\frac{\omega_{log}}{1.2}{\rm exp}[\frac{-1.04(1+\lambda)}{\lambda(1-0.62\mu^*)-\mu^*}],
\end{equation}
where $\mu^*$ is the effective screened Coulomb repulsion constant ($\mu^*$ = 0.13) and $\omega_{\rm log}$ is the logarithmically averaged phonon frequency,

\begin{equation}
\omega_{\rm log}={\rm exp}[\frac{2}{\lambda}\int{\frac{d\omega}{\omega}\alpha^2F(\omega){\rm ln}(\omega)}].
\end{equation}

\section{Results and discussion}

\begin{figure}[tbp]
\centerline{\includegraphics[scale=0.42]{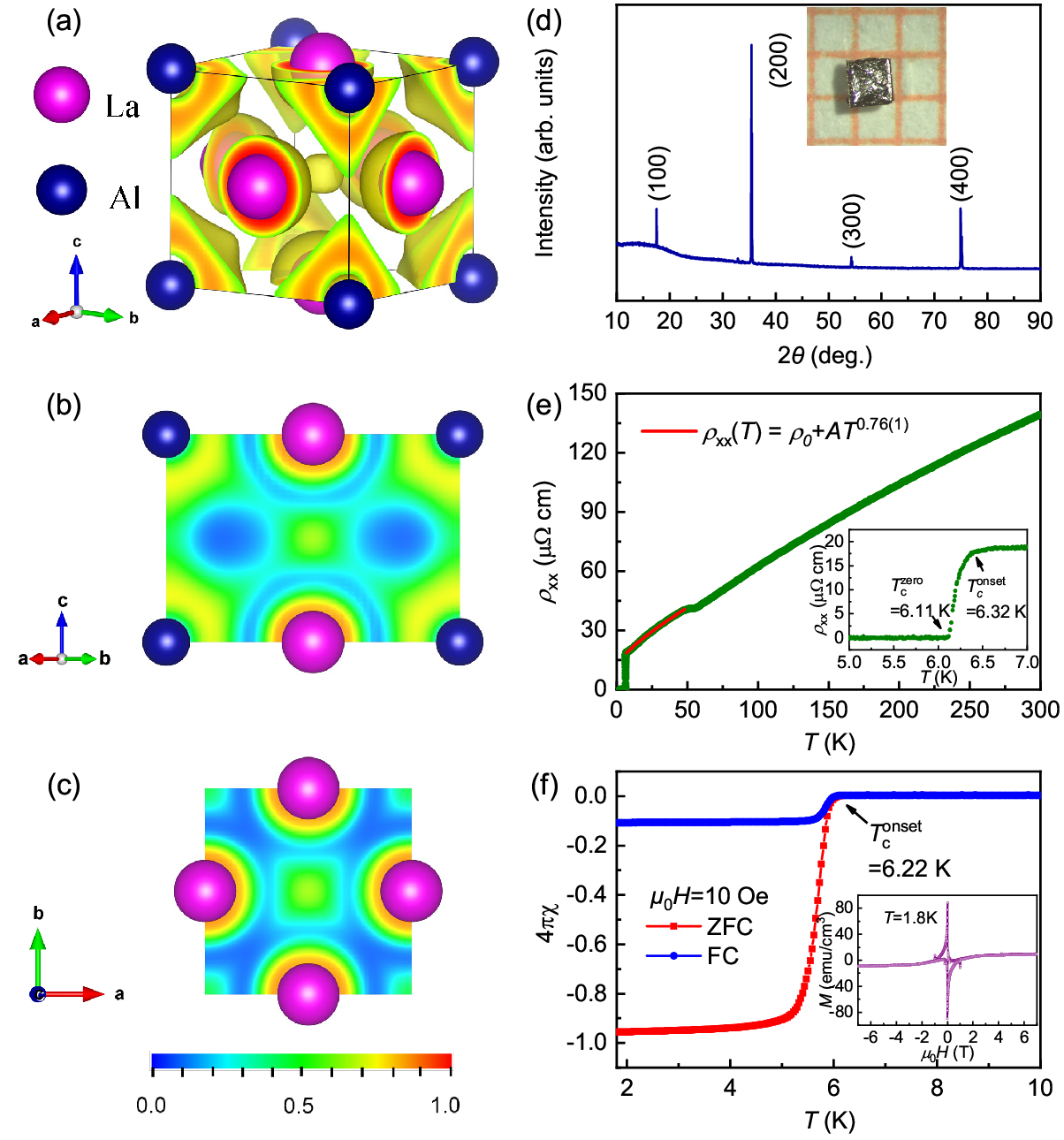}} \vspace*{-0.3cm}
\caption{The electron localization function (ELF) maps for La$_3$Al:
(\textrm{a}) the three-dimensional (3D) map, (\textrm{b}) and (\textrm{c}) the two-dimensional (2D) map projected onto the (110) and (001) plane, respectively. The isosurface values set to 0.5. Pink and blue balls represent La and Al atoms, respectively.
(\textrm{d}) XRD pattern of a La$_{3}$Al single crystal. Inset: photo of typical La$_{3}$Al single crystal on a 1 mm grid paper.
(\textrm{e}) Temperature dependence of $\rho_{xx}(T)$ at zero field for La$_3$Al single crystal. Inset: Enlarged view of $\rho_{xx}(T)$ curve near $T_{c}$.
(\textrm{f}) Temperature dependence of $4\pi \chi(T)$ of La$_3$Al single crystal at $\mu_{0}H=$ 1 mT with ZFC and FC modes. Inset: isothermal magnetization loops at $T=$ 1.8 K.}
\end{figure}

\begin{table*}
\centering
\caption{Crystallographic data and atomic positions for La$_{3}$Al at different temperatures.}
\tabcolsep 0.2in
\renewcommand\arraystretch{1.5}
\begin{tabular}{cccc}
\hline\hline
$T$ (K) & 40 & 100 & 300   \\
space group & $Pm$-3$m$ & $Pm$-3$m$ & $Pm$-3$m$  \\
crystal system & cubic & cubic & cubic  \\
$a$ ({\AA}) & 5.0489(1) & 5.0531(2) & 5.0589(1) \\
$V$ ({\AA}$^{3}$) & 128.704(8) & 129.025(15) & 129.470(8) \\
$Z$    & 1 & 1 & 1 \\
dimens min/mid/max(mm$^{3}$) & 0.04/0.06/0.10 & 0.04/0.06/0.10 &  0.04/0.06/0.10 \\
calcd density (g cm$^{-3}$) & 5.725 & 5.710 & 5.691 \\
abs coeff (mm$^{-1}$)& 24.420 & 24.359 & 24.275 \\
$h$ & -6 $\leq$ $h$ $\leq$ 6 & -7 $\leq$ $h$ $\leq$ 7 & -6 $\leq$ $h$ $\leq$ 6  \\
$k$ &  -6 $\leq$ $k$ $\leq$ 6 & -7 $\leq$ $k$ $\leq$ 7 & -6 $\leq$ $k$ $\leq$ 6  \\
$l$ &  -6 $\leq$ $l$ $\leq$ 6 & -7 $\leq$ $l$ $\leq$ 5 & -5 $\leq$ $l$ $\leq$ 6  \\
reflns collected/unique/$R$(int)& 2383/53/0.0689 & 1633/63/0.0516 & 1550/53/0.0552  \\
data/params/restraints  & 53/5/0 &  63/5/0 & 53/5/0  \\
GOF on $F^{2}$ & 1.340 & 1.304 & 1.299  \\
$R$ indices (all data) ($R$1/$wR$2)$^{a}$ & 0.0472/0.0962 & 0.0441/0.0904 & 0.0486/0.0982  \\
\hline
atom & La/Al & La/Al & La/Al \\
site    & 3$c$/1$a$ & 3$c$/1$a$ & 3$c$/1$a$ \\
$x/a$ & 0/0     & 0/0     & 0/0\\
$y/b$ & 0.5/0 & 0.5/0  & 0.5/0  \\
$z/c$ & 0.5/0  & 0.5/0  & 0.5/0  \\
$U_{\rm eq}$ (A$^{2}$) & 0.0041(12)/0.007(4) & 0.0046(8)/0.006(3) & 0.0107(12)/0.013(5) \\
\hline\hline
\end{tabular}
\end{table*}

As shown in Fig. 1(a), for cubic La$_3$Al, La atoms are located at the face-centered positions of the cubic lattice and Al atoms occupy the vertex positions.
The fit of single crystal XRD measured at room temperature confirms that the grown La$_3$Al single crystal is cubic phase with lattice parameter $a=$ 5.0589(1) \AA\ (Table I).
According to the stoichiometric ratio of atomic species in La$_3$Al, we deduce the existence of excess electrons in this material. Figs. 1(a)-(c) show the three-dimensional (3D) map for the electron localization functions (ELF) as well as the two-dimensional (2D) maps projected onto the (001) and (110) planes.
As can be seen clearly, there are indeed partial electrons separated from the nuclei and confined at the lattice cavites (center of empty La$_6$ octahedra), so called IAEs.\cite{Dye2,Matsuishi,IAEs2,IAEs4} Based on the Bader charge analysis, each IAE carries a charge quantity of 0.53e$^{-}$, functioning as a non-nuclear attractor with the capacity to attract electronegative atoms.\cite{electronegative1,electronegative2} Hence, La$_3$Al can be classified in the category of the well-known electride compounds.\cite{Dye1,electride1}
In addition, the XRD pattern of a La$_3$Al single crystal shows that all peaks can be indexed well by the ($h$00) diffraction indices of cubic structure (Fig. 1(d)), indicating that the surface of crystal is normal to the $a$-axis. The cubic shape of the La$_3$Al crystal (inset of Fig. 1(d)) is consistent with the XRD pattern and its cubic crystallographic symmetry.

The temperature dependent zero-field electrical resistivity $\rho_{xx}(T)$ of La$_{3}$Al shows a good metallic behavior and the residual resistivity ratio (RRR), defined as $\rho$(300 K)/$\rho$(6.5 K), is about 8.2 (Fig. 1(e)).
Interestingly, when $T$ decreases to $\sim$ 55 K, the $\rho_{xx}(T)$ curve exhibits an anomalous kink. Similar anomaly has also been observed occasionally in hexagonal La$_{3}$Al.\cite{Garde, Chen, H-La3Al}
However, the result of single-crystal XRD measured at $T=$ 40 K indicates that there is no structural transition (Table I).
Thus, this anomaly may not originate from structural transition or such structural transition is too subtle to be resolved by ordinary XRD. Such resistivity anomaly is worthy of studying in the future.
The  $\rho_{xx}(T)$ curve between 7 K and 48 K  can be fitted well by using the formula $\rho_{xx}(T)=\rho_{0}+AT^{n}$ and the obtained $n$ is 0.76(1), which is significantly different from the values due to conventional electron-phonon ($n=$ 5) or electron-electron scattering ($n=$ 2).
With lowering $T$ further, there is a sharp drop of resistivity at $T_{c}^{\rm{onset}}=$ 6.32 K and the $\rho_{xx}(T)$ becomes zero at $T_{c}^{\rm{zero}}=$ 6.11 K, which is caused by superconducting transition (inset of Fig. 1(e)).
The $T_{c}$ of cubic La$_{3}$Al is close to that of hexagonal one ($T_{c}=$ 5.80 - 6.37 K).\cite{Garde, H-La3Al}
Fig. 1(f) shows the magnetic susceptibility $4\pi\chi(T)$ curves measured at 1 mT with zero-field cooling (ZFC) and field-coolling (FC) modes. Both curves exhibit a diamagnetic signal at $T_{c}^{\rm{onset}}\sim$ 6.22 K, consistent with the $T_{c}$ value obtained from the $\rho_{xx}(T)$ curve.
Moreover, the value of ZFC 4$\pi\chi(T)$ approaches -1 after considering demagnetization correction, clearly indicating the bulk superconductivity of cubic La$_{3}$Al and excluding the possible contamination of hexagonal phase undoubtedly.
The inset of Fig. 1(f) displays the $M(\mu_0H)$ curve measured at 1.8 K, which shows an obvious hysteresis. It suggests that the cubic La$_{3}$Al is a type-II superconductor, consistent with the bifurcation of ZFC and FC 4$\pi\chi(T)$ curves below $T_{c}$.

\begin{figure}[tbp]
\centerline{\includegraphics[scale=0.35]{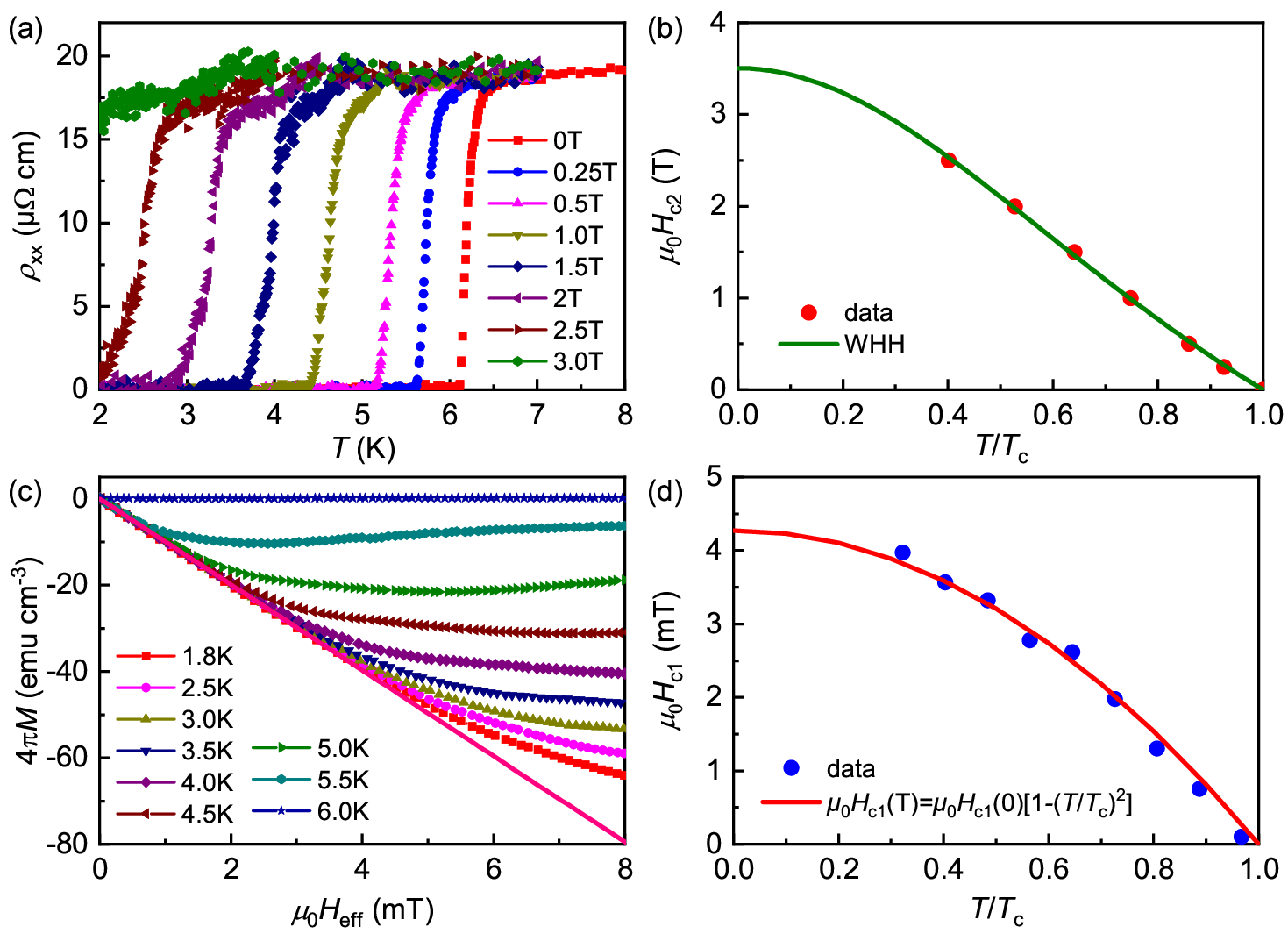}} \vspace*{-0.3cm}
\caption{(\textrm{a}) $\rho_{xx}(T)$ as a function of temperature at various magnetic fields up to 3 T.
 (\textrm{b}) Temperature dependence of $\mu_0 H_{c2}(T)$. The green line represents the fit using the WHH formula.
(\textrm{c}) Low-field parts of $4\pi M(\mu_0H_{\rm eff})$ curves at various temperatures below $T_{c}$. The pink line is the Meissner line.
(\textrm{d}) Temperature dependence of $\mu_0 H_{c1}(T)$. The red line is the fit using the formula $\mu_0 H_{c1}(T)=\mu_0 H_{c1}(0)[1-(T/T_c)^2]$.
}
\end{figure}

Fig. 2(a) shows the temperature dependence of $\rho_{xx}(T)$ at various magnetic fields up to 3 T. With increasing magnetic fields, the superconducting transition becomes broadening and the $T_{c}^{\rm{onset}}$ shifts to lower temperatures gradually.
When $\mu_0H=$ 3 T, the superconducting transition cannot be observed above 2 K.
The upper critical field $\mu_0 H_{c2}(T)$ is determined by the criterion of 50 $\%$ of the normal state resistivity just above superconducting transition, which is shown in Fig. 2(b).
It can be seen that the  $\mu_0 H_{c2}(T)$ increases with decreasing temperature with a slope $\frac{d\mu_0 H_{c2}}{dT}|_{T=T_{c}^{\rm{onset}}(0)}$= -0.57 T K$^{-1}$, where $T_{c}^{\rm{onset}}(0)$ is the superconducting transition temperature at zero field.
The $\mu_0 H_{c2}(T)$ can be fitted well using the Werthamer-Helfand-Hohenberg (WHH) model,\cite{WHH} and the obtained value of $\mu_0 H_{c2}(0)$ is 3.5(1) T. This value is smaller than that of hexagonal La$_3$Al ($\mu_0 H_{c2}(0)=$ 6.95 T).\cite{H-La3Al}
Since the Pauli limiting field $\mu_0$$H_{c2}^P(0)=$ 1.84$T_{c}=$ 10.96 T,\cite{PauliLimit} much larger than $\mu_0 H_{c2}(0)$, the orbital depairing mechanism should be dominant in La$_3$Al.
Using the fitted $\mu_0 H_{c2}(0)$, the calculated Ginzburg-Landau coherence lengths $\xi_{\rm GL}(0)$ from the equation $\xi_{\rm GL} = \sqrt{\Phi_{0}/2\pi\mu_{0}H_{c2}}$ ($\Phi_0=h/2e$ is quantum flux) is 97.1(4) \AA.

Fig. 2(c) shows the low-field $M(\mu_0H_{\mathrm{eff}})$ curves at various temperatures below $T_{c}$.
The $\mu_0 H_{\mathrm{eff}}$ is calculated with considering demagnetization effect using the formula $\mu_0 H_{\mathrm{eff}}=\mu_0 H_{a}-NM$, where $N$ is the demagnetization factor and $\mu_0 H_{a}$ is the external field.\cite{Aharoni}
The estimated demagnetization factor is 0.38 and the fitted slope of $M(\mu_0H_{\mathrm{eff}})$ curve at 1.8 K is -0.993(2), very close to -1 ($4\pi M=-\mu_0 H_{\mathrm{eff}}$).
Thus the full Meissner shielding effect in our measurement provides a reliable way to determine the value of $\mu_0 H_{c1}$.
The $\mu_0 H_{c1}$ is determined as the point deviating from linearity based on the criterion $\Delta 4\pi M=(4\pi M_{m}-4\pi M_{th})=1\times 10^{-6}$ emu cm$^{-3}$, where $4\pi M_{m}$ is the measured moment value and $4\pi M_{th}$ is the calculated moment value at the same field.
The extracted $\mu_0 H_{c1}(T)$ at the different temperatures is shown in Fig. 2(d). The $\mu_0 H_{c1}(T)$ can be fitted well using the formula $\mu_0 H_{c1}(T)=\mu_0 H_{c1}(0)[1-(T/T_{c})^{2}]$ (red solid line) and the obtained $\mu_0 H_{c1}(0)$ is 4.3(1) mT.
Similar to $\mu_0 H_{c2}(0)$, the value of $\mu_0 H_{c1}(0)$ for cubic La$_{3}$Al is also smaller than that of hexagonal La$_3$Al ($\mu_0 H_{c1}(0)=$ 22.17 mT).\cite{H-La3Al}
According to the equation $\mu_0 H_{c1}=\frac{\Phi_0}{4\pi\lambda_{\rm GL}^2}{\rm ln}\frac{\lambda_{\rm GL}}{\xi_{\rm GL}}$, the value of superconducting penetration depth  $\lambda_{\rm GL}(0)$ is 3738(53) \AA.
Correspondingly, the determined Ginzburg-Landau constant $\kappa_{\rm GL} (= \lambda_{\rm GL}/\xi_{\rm GL}$) is 38.5(3), further confirming La$_{3}$Al is a type-II superconductor.

\begin{figure}[tbp]
\centerline{\includegraphics[scale=0.35]{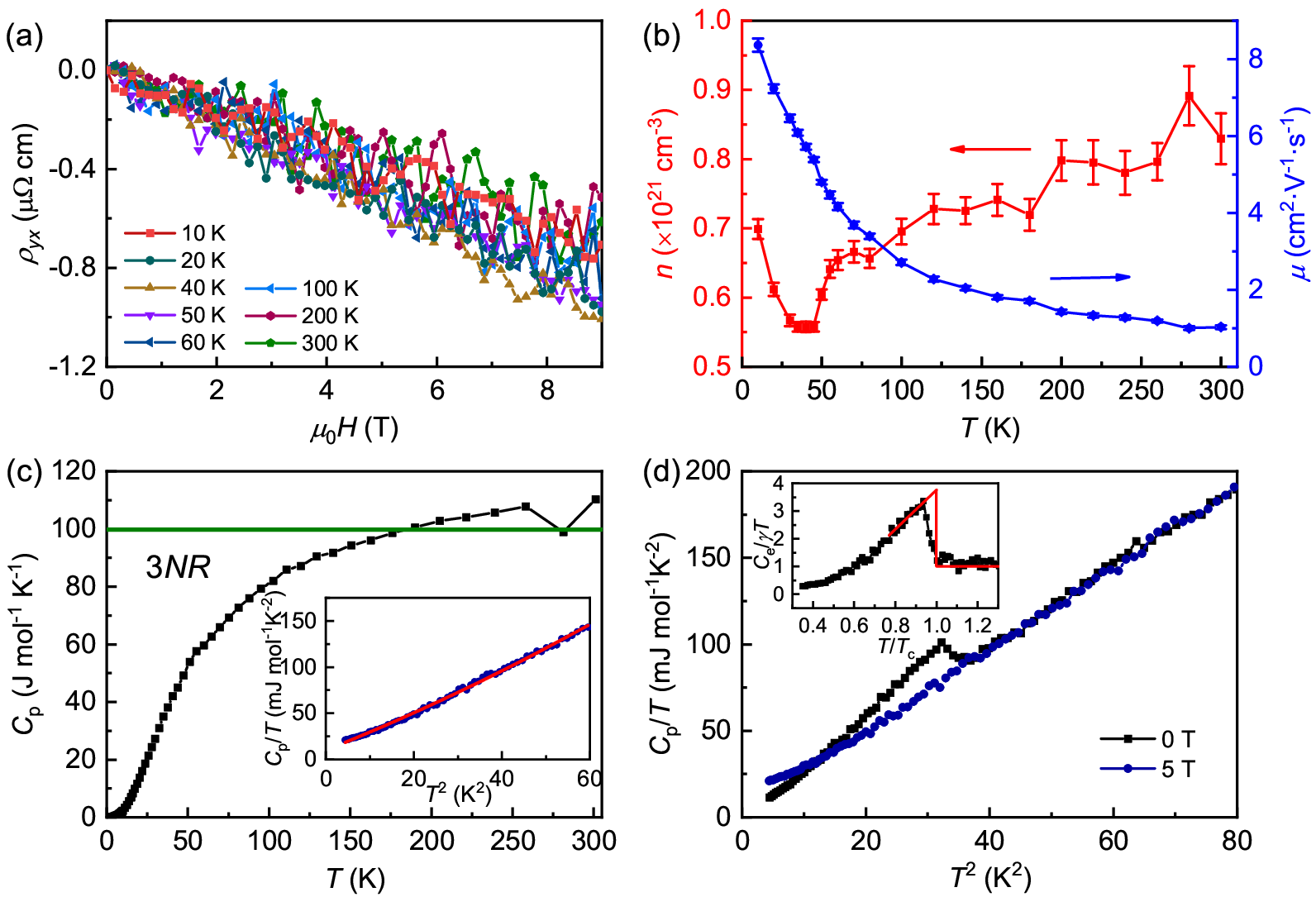}} \vspace*{-0.3cm}
\caption{(\textrm{a}) Field dependence of $\rho_{yx}(\mu _{0}H)$ at various temperatures.
(\textrm{b}) Temperature dependence of derived $n(T)$ and $\mu(T)$.
(\textrm{c}) Temperature dependence of $C_{\rm p}$ from 2 K to 300 K at zero field. Inset: $C_{\rm p}/T$ vs. $T^{2}$ at 5 T. The red solid line represents the fit using the formula $ C_{\rm p}/T=\gamma +\beta T^{2} +\delta T^{4}$.
(\textrm{d}) Low-temperature specific heat $C_{\rm p}/T$ vs. $T^{2}$ at zero field and 5 T. The Inset shows the relationship between $C_{\rm e}/\gamma T$ and $T/T_{c}$.
}
\end{figure}

Fig. 3(a) shows field dependence of Hall resistivity $\rho_{yx}(\mu_0 H)$ at various temperatures.
All curves display a negative slope, suggesting that the electron-type carriers play a dominant role in the transport of La$_{3}$Al.
By using the linear fits of $\rho_{yx}(\mu_0 H)$ curves and combining the result of $\rho_{xx}(T)$ at zero field, the carrier concentration $n(T)$ (red squares) and carrier mobility $\mu(T)$ (blue circles) as functions of temperature can be obtained.
As depicted in Fig. 3(b), with decreasing temperature from 300 K to 60 K, the $n(T)$ decreases slightly from about 0.83(4) to 0.65(1) $\times$10$^{22}$ cm$^{-3}$. With lowering temperature further, it exhibits a sudden drop around 55 K, which is in line with the anomaly at $\rho_{xx}(T)$ curve. At lower temperature, the $n(T)$ increases to 0.70(2)$\times$10$^{22}$ cm$^{-3}$ at 10 K.
Meanwhile, the $\mu(T)$ exhibits a gradual increase from 1.0(1) cm$^2$ V$^{-1}$ s$^{-1}$ at 300 K to 8.4(2) cm$^2$ V$^{-1}$ s$^{-1}$ at 10 K and there is no obvious anomaly at $T\sim$ 55 K.
Fig. 3(c) shows the specific heat $C_{\rm p}(T)$ of La$_{3}$Al measured from 2 K to 300 K at zero field. The value of $C_{\rm p}(T)$ at 300 K approaches the classical value of 3$NR$ ($\sim $ 99.77 J mol$^{-1}$ K$^{-1}$) as the Dulong-Petit law predicts (green solid line), where $N$ (= 4) is the atomic number per formula and $R$ (= 8.314 J mol$^{-1}$ K$^{-1}$) is the ideal gas constant. In addition, there is a small kink near $T=$ 55 K, which might be associated with the resistivity anomaly.
Fig. 3(d) shows the relationship between $C_{p}/T$ and $T^{2}$ at low-temperature region with $\mu_0H=$ 0 T and 5 T. It can be seen that there is a jump at $T_{c}\sim$ 6.06 K for the curve measured at zero field, confirming the bulk superconducting transition in cubic La$_{3}$Al. The $T_{c}$ is in agreement with the values obtained from the resistivity and magnetization measurements.
On the other hand, the field of 5 T suppresses the superconducting transition completely at $T>$ 2 K.
Moreover, the $C_{\rm P}/T$ curve at 5 T can be fitted using the formula $C_{\rm P}(T)/T=\gamma +\beta T^{2} +\delta T^{4}$ (inset of Fig. 3(c)), where the $\gamma$ is electronic specific heat coefficient and the $\beta$ and $\delta$ are the lattice specific heat coefficients.
The fitted $\gamma$ is 9.9(6) mJ mol$^{-1}$ K$^{-2}$ when the values of $\beta$ and $\delta$ are 1.9(1) mJ mol$^{-1}$ K$^{-4}$ and 0.006(1) mJ mol$^{-1}$ K$^{-6}$.
Correspondingly, the calculated Debye temperature $\Theta _{\rm D}$ is 160(1) K using the equation $\Theta_{\rm D}=(12\pi ^{4}NR/5\beta )^{1/3}$.
The inset of Fig. 3(d) plots the $C_{e}/\gamma T$ vs. $T/T_{c}$, where $C_{e}$ is the electronic specific heat which is obtained by subtracting the lattice contribution from the total specific heat.
The extracted specific heat jump $\Delta C_{e}/\gamma T$ at $T_{c}$ is about 2.20, which is larger than the weak coupling value of BCS superconductor (1.43). It indicates that La$_{3}$Al exhibits an superconductivity with intermediate coupling strength.
On the other hand, the electron-phonon coupling constant $ \lambda _{e-ph}$ is obtained from the McMillan equation,\cite{McMillan}

\begin{equation}
\lambda _{e-ph}=\frac{\mu ^{\ast }\ln(1.45T_{c}/\Theta _{D})-1.04}{1.04+\ln(1.45T_{c}/\Theta _{D})(1-0.62\mu ^{\ast })}
\end{equation}

When assuming the Coulomb pseudopotential $\mu ^{\ast}\approx $ 0.13, the value of $\lambda _{e-ph}$ is determined to be 0.88(1) by using $T_{c}=$ 6.3 K and $\Theta _{\rm D}=$ 160(1) K, also confirmed the intermediately coupled BCS superconductivity of La$_{3}$Al.\cite{Allen}

\begin{figure}[tbp]
\centerline{\includegraphics[scale=0.5]{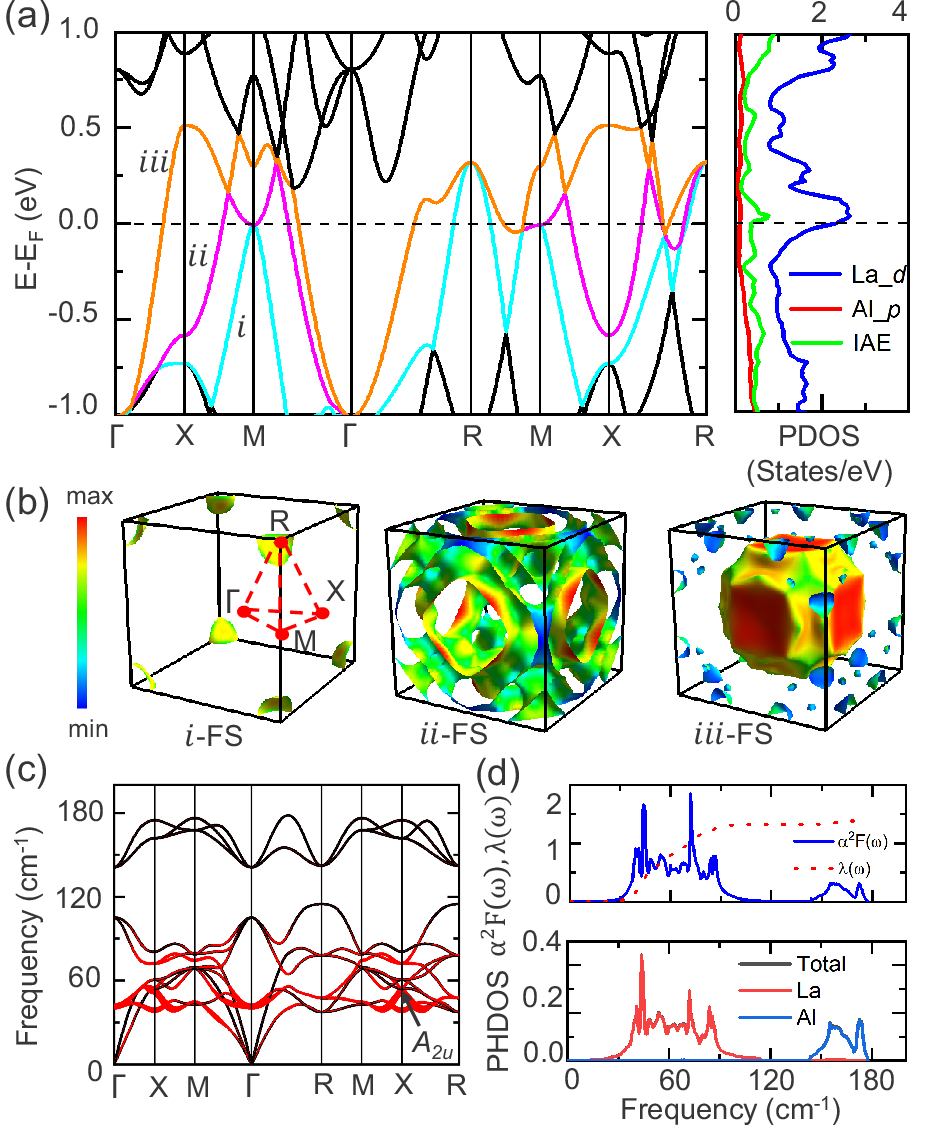}}
\caption{(\textrm{a}) The orbital-resolved electronic band structure and PDOS of La$_3$Al.
(\textrm{b}) The Fermi-velocity-projected FS sheets corresponding to the bands \textit{i}, \textit{ii}, and \textit{iii} in (\textrm{a}) , respectively. (\textrm{c}) Phonon dispersion curves. The size of red dots represents the electron-phonon coupling (EPC) strength $\lambda_{{\bf q}\nu}$. (\textrm{d}) The Eliashberg spectral function $\alpha^2F(\omega)$,  frequency-dependent EPC constant $\lambda(\omega)$, the phonon density of states (PHDOS).}
\end{figure}

Fig. 4(a) shows the electronic band structure and the partial density of states (PDOS) of La$_3$Al calculated without the spin-orbital coupling (SOC).
There are three bands, labeled \textit{i}, \textit{ii}, and \textit{iii}, crossing the Fermi level $E_{\rm F}$, among which there are short flat bands along the R-M path of the BZ.
Based on the orbital analysis, these short flat bands are dominated by the La 5$d$ electrons and the IAEs (Fig. S1 in Supplemental Material (SM)).
According to the PDOS in Fig. 4(a), we find that the electronic states near $E_{\rm F}$ are mainly contributed by the La 5$d$ orbitals and the IAEs, which are consistent with the band features.
More importantly, there is a van Hove singularity around $E_F$, which results in a large DOS of 1.60 states/eV per atom at $E_{\rm F}$. The Fermi surface (FS) sheets for the bands \textit{i}, \textit{ii}, and \textit{iii} are correspondingly shown in Fig. 4(b), on which the Fermi velocities are displayed with the color scales, where the red and blue colors represent the highest and zero Fermi velocities, respectively. It is clear that the two FS sheets of bands \textit{ii} and \textit{iii} have a lower Fermi velocity around the M point, which is favorable for strong EPC.\cite{lower-velocity1,lower-velocity2} Besides, we also examined the electronic structure with the inclusion of SOC, but found that both the energy dispersion and the total DOS near $E_F$ are almost unchanged (Fig. S2 in SM).

To better understand the superconducting properties of La$_3$Al, we subsequently performed the EPC calculations. The calculated total EPC constant $\lambda$ is 1.29, which is slightly larger than experimental value. These lead to a superconducting $T_c$ of 7.37 K based on the McMillan-Allen-Dynes formula (Eq. (\ref{eq5})).
After further consideration of SOC, $\lambda$ is reduced to 1.28 and logarithmic phonon frequency increases from 53.85 cm$^{-1}$ to 58.52 cm$^{-1}$. The decrease in $\lambda$ and the increase in $\omega_{\rm log}$ leads to a slight change in $T_c$ to 6.78 K, which is in good agreement with the measured one (Fig. 1(d)).
From the momentum- and mode-resolved EPC parameter $\lambda_{{\bf q}\nu}$ (Fig. 4(c)), we learn that the largest contribution comes from the acoustic branches around X point and $\Gamma$ point, which results in two high peaks (around 44 cm$^{-1}$ and 72 cm$^{-1}$) in the Eliashberg spectral function $\alpha^2F(\omega)$ (the top part of Fig. 4(d)).
Combined the frequency-dependent EPC parameters $\lambda(\omega)$ and the phonon density of states (PHDOS) (Fig. 4(d)), we can see that the La vibrations play a dominant role in the superconductivity. Specifically, the strongest EPC located at the softened acoustic branch at $X$ point and is related to the $A_{2u}$ mode (Fig. 4(c)). This mode corresponds to the vibrations of La atoms in the La-Al layers along the $c$ axis (Fig. S3(a) in SM). Meanwhile, we found that when La atoms have small displacements along the normal mode coordinates of the $A_{2u}$ mode, there are noticeable shifts in the electronic bands around the Fermi level at the $M$ point (Fig. S3(b) in SM), whose orbital weights derive from the IAEs and La 5$d$ electrons (Fig. S1 in SM). These electronic band shifts indicate that the $A_{2u}$ phonon strongly couples with those electronic states.
In short, we propose that the superconducting pairing in La$_3$Al belongs to the conventional BCS type and the superconductivity originates from the coupling of the electronic states of La 5$d$ and IAEs with the La-derived low-frequency phonons.

\section{Conclusion}

In summary, the cubic La$_{3}$Al single crystals are grown successfully using La flux and the  physicals properties as well as electronic structures are investigated in detail.
Bulk superconductivity with $T_{c}^{\rm{onset}}=$ 6.32 K has been observed from electrical resistivity, magnetization and specific heat measurements.
Further analysis reveals that cubic La$_{3}$Al is an intermediately coupled type-II BCS superconductors. Moreover,  cubic La$_{3}$Al can host IAEs and thus can be considered as an electride superconductor, similar to cubic La$_{3}$In. Such study will deepen our understanding of electride superconducting materials.

\section{Acknowledgments}

This work is supported by the National Key R\&D Program of China (Grant Nos. 2022YFA1403800, 2023YFA1406500 and 2022YFA1403103), the National Natural Science Foundation of China (Grant Nos. 12274459, 12174443, and 12074013).  Computational resources were provided by the Physical Laboratory of High-Performance Computing at Renmin University of China.

$^{\dag}$ These authors contributed to this work equally.

$^{\ast}$ Corresponding authors:  K. Liu (kliu@ruc.edu.cn) and H. C. Lei (hlei@ruc.edu.cn).

\end{document}